\newif\ifPDF
\ifx\pdfoutput\undifine\PDFfalse
\else\ifnum\pdfoutput >0\PDFtrue
\else\PDFfalse
\fi
\fi

\ifPDF
   \documentclass[10pt,twocolumn,twoside,a4paper,pdftex]{article}
\else
   \documentclass[10pt,twocolumn,twoside,a4paper,dvips]{article}
\fi

\usepackage[small]{caption}
\def\abstract{\begin{center}
{\it \large Abstract\vspace{-.5em}\vspace{0pt}}
\end{center}}

\usepackage{amsfonts}
\usepackage{amssymb}
\usepackage{verbatim}
\usepackage{graphicx}
\usepackage{hyperref}

\ifPDF
   \hypersetup{%
     pdftoolbar=treu,%
     pdfmenubar=true}
\else
   \usepackage{epsfig}
   \usepackage{pslatex}
\fi

\newcommand{\ssection}[1]{%
     \section[#1]{\centering \sc #1}}
\newcommand{\ssubsection}[1]{%
     \subsection[#1]{ \centering \it #1}}

\def\abstract{\begin{center}
{\it \large Abstract\vspace{-.5em}\vspace{0pt}}
\end{center}}

\title{
\vspace{-1cm} 
\Large {The Baikal Neutrino Telescope: Results, Plans, Lessons}}
\author{\large C. Spiering for the BAIKAL Collaboration
\\ \normalsize DESY Zeuthen, Platanenallee 6, D-15738 Zeuthen, Germany\\
\normalsize christian.spiering@desy.de}

\begin{document}
\date{}
\maketitle

\thispagestyle{empty}
\pagestyle{empty}

\noindent
\begin{center}
{\it The BAIKAL Collaboration:\\}
\end{center}
{\small
V.~Aynutdinov$^1$,
V.~Balkanov$^1$,
I.~Belolaptikov$^4$,
L.~Bezru\-kov$^1$,
N.~Budnev$^2$,
A.~Chensky$^2$,
D.~Chernov$^3$,
I.~Danil\-chen\-ko$^1$,
Zh.-A.~Dzhil\-kibaev$^1$,
G.~Domo\-gatsky$^1$,
A.~Dyachok$^2$,
O.~Gapo\-nenko$^1$,
O.~Gress$^2$,
T.~Gress$^2$,
A.~Klabukov$^1$,
A.~Klimov$^8$,
S.~Klimu\-shin$^1$,
K.~Konischev$^4$,
A.~Koshech\-kin$^1$,
V.~Kulepov$^6$,
L.~Kuzmi\-chev$^3$,
Vy.~Kuznetzov$^1$,
B.~Lubsan\-dor\-zhiev$^1$,
S.~Mikheyev$^1$,
M.~Mile\-nin$^6$,
R.~Mirga\-zov$^2$,
N.~Moseiko$^3$,
E.~Osipova$^3$,
A.~Panfi\-lov$^1$,
G.~Pan'kov$^2$,
L.~Pan'kov$^2$,
Yu.~Parfenov$^2$,
A.~Pavlov$^2$,
E.~Plis\-kov\-sky$^4$,
P.~Pokhil$^1$,
V.~Polecshuk$^1$,
E.~Popova$^3$,
V.~Prosin$^3$,
M.~Rosa\-nov$^7$,
V.~Rubtzov$^2$,
Yu.~Semeney$^2$,
B.~Shaibo\-nov$^1$,
C.~Spiering$^5$,
B.~Tara\-shansky$^2$,
R.~Vasiliev$^4$,
E.~Vyatchin$^1$,
R.~Wisch\-newski$^5$,
I.~Yashin$^3$,
V.~Zhukov$^1$ \\ 
{\it (1) Institute for Nuclear Research, Moscow, Russia, 
(2) Irkutsk State University, Irkutsk,Russia, 
(3) Skobeltsyn Institute of Nuclear Physics  MSU, Moscow, Russia,
(4) Joint Institute for Nuclear Research, Dubna, Russia,
(5) DESY--Zeuthen, Zeuthen, Germany,
(6) Nizhni Novgorod State Technical University,
(7) St.Peterburg State Marine University, St.Peterburg, Russia,
(8) Kurchatov Institute, Moscow, Russia} \\
}

\begin{abstract}
We review 
recent results on the search for
high energy extraterrestrial neutrinos,
neutrinos induced by WIMP annihilation and
neutrinos coincident with Gamma Ray Bursts
as obtained with the Baikal neutrino telescope NT-200.
We describe the moderate upgrade of NT-200
towards a $\sim$10 Mton scale detector NT-200$+$.
We finally draw a few lessons from our
experience which may be of use for other
underwater experiments.
\end{abstract}

\ssection{Introduction}

The Baikal Neutrino Telescope NT-200  is operated in Lake 
Baikal, Siberia,  at a depth of \mbox{1.1 km}. 
A description of the detector as well
as physics results from data collected in 1996 and 1998
(70 and 234 live days, respectively) 
have been presented elsewhere \cite{96-98-a,96-98-b}. 
Here we present
new limits including data taken in 1999 (268 live days).

We note that the year 2003 marks the
tenth anniversary of the deployment of NT-36, the 
pioneering first
stationary underwater array \cite{NT-36,CC}. 
Fig.1 shows a textbook
neutrino event (an upward moving muon track)
recorded with the early 4-string configuration of 1996 \cite{APP2}.

\begin{figure}[ht]
\begin{center}
\includegraphics[width=.35\textwidth]{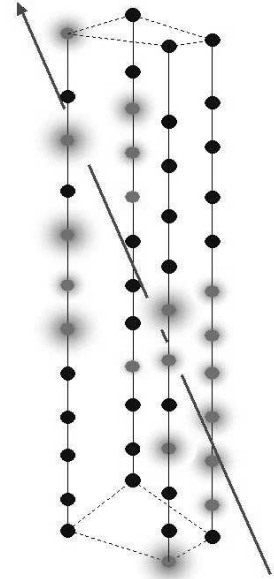}
\end{center}
\vspace{-0.5cm}
\caption{
Upward going muon track with 19 fired
channels, recorded in 1996. Hit channels are
marked in color, with the size of the aura indicating
the recorded amplitude.
}
\label{goldplated}
\end{figure}

\ssection{Search for extraterrestrial high energy neutrinos}

The survey for high energy neutrinos is focused to
bright cascades produced at the neutrino interaction
vertex in a large volume around the neutrino telescope.
Lack of significant light scattering allows to monitor a 
volume exceeding the geometrical volume by an order of magnitude.
This results in sensitivities of NT-200 comparable
to those of the much larger Amanda-B10 detector. 
The background to this search are bright bremsstrahlung flashes
along muons passing far outside the array
(see \cite{diffuse} for details).

Candidate events do not show a statistically
significant excess of hit multiplicites compared to the
simulated background from atmospheric muons.
Assuming an $E^{-2}$ shape of the neutrino spectrum and a
flavor ratio  $\nu_e:\nu_{\mu}:\nu_{\tau}=1:1:1$,  
the new, preliminary 90\% c.l. upper limit with respect
to the flux of all three flavors is
$\Phi_{\nu_e +  \nu_{\mu} + \nu_{\tau} }E^2=1.3\cdot10^{-6} 
\mbox{cm}^{-2}\mbox{s}^{-1}\mbox{sr}^{-1}\mbox{GeV}$,
about twice below previous results \cite{diffuse}.
The preliminary limit on $\tilde{\nu_e}$ at the W - resonance 
energy is:
$\Phi_{\tilde{\nu_e}} \leq 5.4 \times 
10^{-20} 
\mbox{cm}^{-2}\mbox{s}^{-1}\mbox{sr}^{-1}\mbox{GeV}^{-1}.$
Fig.~2  shows the experimental upper limits obtained by 
BAIKAL (this work), and AMANDA \cite{Amanda}, as well as
the projected sensitivity for NT-200+ (see below).
Limits are compared to SSDS \cite{SSDS} and MPR \cite{MPR}
predictions.
The slanted lines at the left side
represent the atmospheric neutrino
fluxes (dashed for $\nu_{\mu}$, solid for $\nu_e$).

\vspace{-2mm}
\begin{figure}[h]
{\hspace{-2mm}
\includegraphics[width=.37\textwidth,angle=-90]{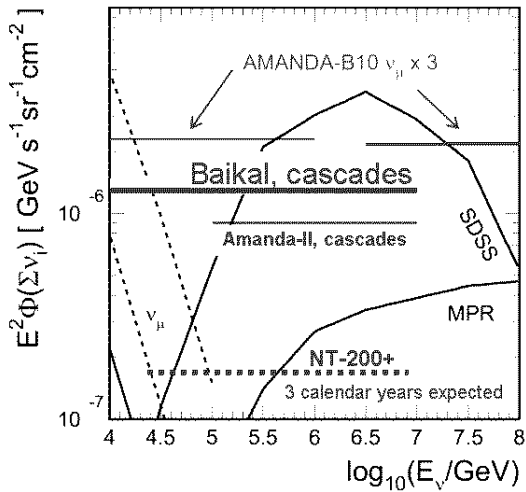}}
\vspace{-2mm}
\caption{
Experimental upper limits on neutrino fluxes
(see text for explanation).
}
\label{diffuse}
\end{figure}

\ssection{Search for neutrinos coincident with GRBs}

We have searched for high energy cascades coincident with
772 Gamma Ray Bursts (GRB) recorded between April~1998 and 
February~2000  by the BATSE detector and falling into on-line
periods of NT-200 (386 triggered GRB, 336 non-triggered from
Stern catalogue). After cuts for background reduction, we 
are left with one event within one of the
$\pm$100-second windows, where 0.47 events would have been
expected from accidental coincidences. Precise 
limits on the fluence are being derived at present. 
With an effective volume in the Megaton range, NT-200
is the largest Northern detector for high energy neutrinos
from GRBs.

\vspace{-2mm}

\ssection{Search for neutrinos from WIMP annihilation}

The search for WIMPs with the Baikal
neutrino telescope is based on a possible signal of
nearly vertically upward going muons, exceeding
the flux of atmospheric neutrinos (see \cite{WIMP}).
Note that the threshold of $\sim$ 10 GeV for this analysis
is lower than that of $\sim$ 15 GeV
for atmospheric neutrinos spread
across the full lower hemisphere.
Fig.~3 (top) demonstrates that the angular distribution
of events passing a special filter for
events close to the vertical is well described
by simulations including the effect of 
neutrino oscillations 
(assuming $\delta m^2 = 2.5 \cdot 10^{-3}$ eV$^2$).

\begin{figure}[h]
\begin{center}
\includegraphics[width=.34\textwidth]{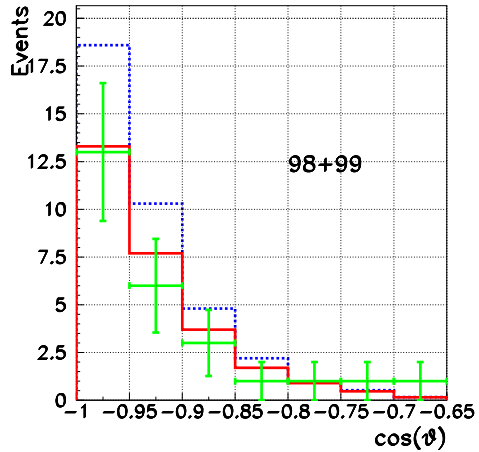}
\vspace{-0.4cm}
\end{center}
\caption{
Angular distribution of selected neutrino candidates 
compared to expected distributions including (full line)
and excluding (dotted line)
the effect of oscillations.
}
\end{figure}

With no
significant excess observed, we derive improved 
upper limits on the flux of muons from the direction
of the center of Earth related to WIMP annihilation.
Fig.~3 (bottom) compares our new limits to those obtained by other 
experiments (see \cite{WIMP} for references).

\begin{figure}[ht]
\begin{center}
\includegraphics[width=.35\textwidth,angle=-90]{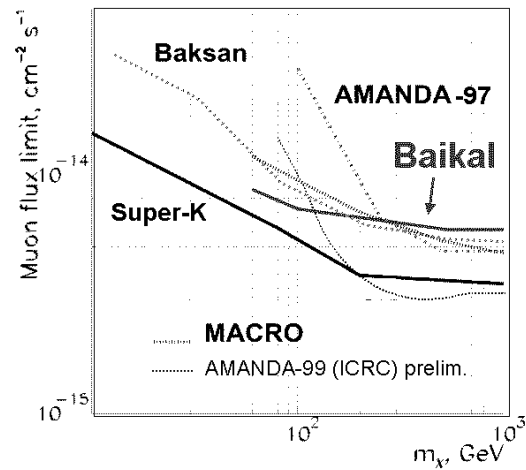}
\end{center}
\vspace{-0.2cm}
\caption{ Flux limits from different experiments
as a function of WIMP mass.
}
\end{figure}

\ssection{Upgrade to NT200+}

NT-200$+$ is an upgrade of NT-200 by 
three sparsely instrumented distant outer strings
(see Fig.~5).
The fenced volume is a few dozen Mtons.
A prototype string of 140 m length
with 12 optical modules was deployed in March 2003, and
electronics, data acquisition
and calibration systems for NT-200$+$ have been tested.
The 3 strings 
allow for dramatically better vertex reconstruction
of high energy cascades than with NT-200 alone
(see Fig.~6). This, in turn, allows a much
better determination of the energy and makes
NT-200+ a true discovery detector with
respect to reactions within the mentioned fiducial volume.
(This is in contrast to NT-200 which is an excellent
exclusion experiment but would have certain difficulties
to prove -- for a single event outside its geometrical
volume -- that the signal is indeed due to a far high-energy event
and not to a closer medium-energy event.)
NT-200+ will be installed in 2004 and 2005.

\begin{figure}[h]
\begin{center}
\includegraphics[width=.31\textwidth,angle=-90]{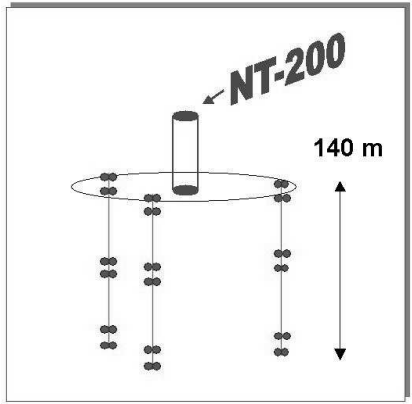}
\end{center}
\caption{
Configuration of NT-200+.
}
\end{figure}

\begin{figure}[ht]
{\hspace{14mm}
\includegraphics[width=.29\textwidth]{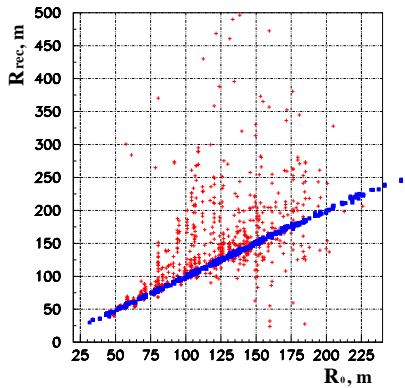}}
\vspace{-0.1cm}
\caption{
Reconstructed vs. simulated coordinates of cascades in NT-200+
(black rectangles) and NT-200 (crosses).
}
\end{figure}

\ssection{A Giga-ton Detector ?}

First discussions have started on an even 
larger detector of the order of a cubic kilometer in Lake
Baikal. The configuration of such a detector
will be consequently tailored to certain
classes of events and should complement Mediterranean
arrays. A possible design was sketched in \cite{96-98-b}.
Given the drawbacks of stronger absorbtion and shallower
depth compared to Mediterranean sites, this idea rests
on the proven, stable deployment procedures,
lower-cost considerations and possible national funding 
strategies. However, even in the case of
strong national funding, such a project would need
international participation.

\ssection{Some lessons}

We finally summarize some lessons which may be of use for
all underwater arrays.

\ssubsection{Leakage}

The early history of Baikal could have been written
as a history of
leaking feed-throughs, connectors and pressure housings.
After having designed a special connector tailored to depths up to
$\sim$ 1.6 km (the maximum depth in Lake Baikal),
this phase was overcome. NT-36 did not suffer from
leaks. In
the mean time, more than 1000 connectors and feed-throughs
are in operation at 1.1 km depth, and the failure rate is 
less than 1 per year. Clearly, detectors much deeper,
like the Mediterranean ones, have to make use of different
technologies. Still, is seems now obvious that instruments
with hundreds of pressure spheres and more than thousand 
connectors can work in deep water over many years, with
tolerable leakage rates -- a statement which ten years
ago would have been questioned by many  oceanographers.

\ssubsection{Sedimentation}

Sedimentation and bio-fouling are a concern in Lake Baikal.
The performance of the up-down symmetric NT-36 detector, 
with the same number of optical modules facing
upward and downward,
was particularly affected by sedimentation \cite{NT-36}.
The sensitivity of
the upward looking modules with respect to atmospheric 
muons decreased to 35\% after one year \cite{OM}. At present,
only two of the 12 layers of NT-200 have upward 
facing modules. The
glass spheres are dressed with a tapering hat of plastic
foil. Much of the sediments glide along the foil and do not
stick to the hat. The effect of sedimentation, i.e.
a progressively decreasing
sensitivity of the upward looking modules, is
weakened, and an acceptable functionality may be kept
by cleaning the hats once every two years. This is important
since the outer strings of NT-200+ will be up-down symmetric. 
We conclude that symmetric configurations  can be 
operated in Lake Baikal, but only if the optical moudules 
can be easily
hauled up, inspected and cleaned. The same limitation
should hold for those Mediterannean sites with strong 
biological activity.

\ssubsection{Reliability and repair}

The possibility to haul up the array and to repair
connectors was a key for NT-200's operation. In the very
beginning, about 40\% of the optical modules turned non-operational
after one year of operation due to failures in
electronics. Clearly this was not tolerable,
be it with or without the possiblity to repair failed components.
In the mean time the failure rate reduced to less than
5\% per year. This lower rate, combined with the possibility to repair
failed components results in a tolerable regime of
operation and maintenance. From our experience we would
recommend designs which give a not too complicated
access to deployed parts of the detector.

\ssubsection{Electro-corrosion}

With the reliability of the detector itself improving from 
year to year, we were faced by another, non-expected problem:
we observed electro-corrosion along the cables to shore.
Actually electro-corrosion was thought to be negligible since
the fresh water of Lake Baikal is nearly free of minerals
and salts and is a bad conductor of electricity. Much of
the return current was thought to flow along the outer metal
jacket of the shore cable (white arrows in figure 7)
and not through the water. 

\begin{figure}[ht]
\begin{center}
\includegraphics[width=.30\textwidth,angle=-90]{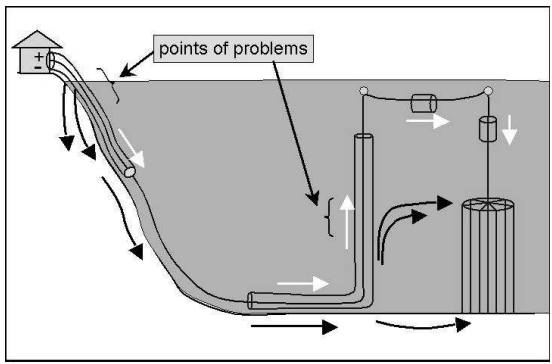}
\vspace{-0.4cm}
\caption{
Electro-corrosion along shore cables. See text for explanations.
}
\end{center}
\end{figure}

During the last year it became
clear that the cable arming showed damages at the points
indicated in the figure. The reason seems to be that
close to the shore the current leaks into ground and
propagates there with a low resistance compared to the cable,
and that also current flows trough the water at the depth
of the big metallic, umbrella-like support frame
of NT-200 (black arrows). At these points,
strong electro-corrosion is observed. It led
to the loss of two of the four shore cables in the last year.
The cables will be replaced by new cables in 2004, and the
distance between the vertical part of the shore cable and
NT-200 will be increased. 

Electro-corrosion should be an even stronger
issue for salt water detectors, as shown by the
DUMAND history. The Baikal experience
can certainly not be translated to different 
designs of power distribution and return current.
Still it may suggest that one should be prepared
to unpleasant surprises and necessary
design changes.

\ssubsection{Staged approach}

A staged approach seems mandatory when building a large
instrument in a new and challenging environment like deep water.
The idea that one can already in the design phase
account for all possible problems
has been disqualified not only by the Baikal
experience but also by DUMAND, NESTOR and ANTARES.
The operation of prototype detectors and the readiness
for technological modifications is of key importance.

\ssubsection{Looking beyond the geometrical volume}

Underwater detectors do not suffer from strong light
scattering as does ice. Therefore the timing information from
distant light sources is not strongly dispersed.
This opens the possibility to monitor
a huge volume around the detector. 
A few strings far outside, like those planned for
NT-200+, can dramatically improve
the sensitivity to PeV processes. Due to
best background rejection, the most efficient region 
is that {\it below} the geometrical volume of the detector.
Apart from stronger bottom water currents and slighlty
worse water quality close to ground, this may be another
argument not to place the detector as deep as possible,
but a few hundred meters above ground.

\vspace{0.7cm}

\noindent
{\it This work was supported by the Russian Ministry of Research,
the German Ministry of Education and Research and the Russian Fund of Basic 
Research ( grants } \mbox{\sf 03-02-31004}, \mbox{\sf 02-02-17031}, 
\mbox{\sf 02-07-90293} {\it and} \mbox{\sf 01-02-17227}),
{\it Grant of President of Russia} \mbox{\sf NSh-1828.2003.2} 
{\it and by the Russian Federal Program ``Integration'' 
(project no.} \mbox{\sf E0248}).

\end{document}